\documentclass[aps,prl,superscriptaddress,amsfonts,amssymb,amsmath,nofootinbib,twocolumn,floatfix]{revtex4-2}
 \usepackage{color,graphicx}
 \usepackage[utf8]{inputenc}
 \usepackage[T2A]{fontenc}
 \usepackage[russian,english]{babel}
 \definecolor{darkblue}{rgb}{0,0,0.7}
\definecolor{darkred}{rgb}{0.7,0,0}
\definecolor{darkgreen}{rgb}{0,0.4,0}
\usepackage[unicode, colorlinks, citecolor=darkblue, linkcolor=darkred, urlcolor=blue]{hyperref}

\allowdisplaybreaks
\begin{document}

\title{Back action evading electro-optical transducer}

\author{Albert I. Nazmiev}
\affiliation{Faculty of Physics, M.V. Lomonosov Moscow State University, Leninskie Gory, Moscow 119991, Russia}

\author{Andrey B. Matsko}

\affiliation{Jet Propulsion Laboratory, California Institute of Technology, 4800 Oak Grove Drive, Pasadena, California 91109-8099, USA}

\author{Sergey P. Vyatchanin}

\affiliation{Faculty of Physics, M.V. Lomonosov Moscow State University, Leninskie Gory, Moscow 119991, Russia}
\affiliation{Quantum Technology Centre, M.V. Lomonosov Moscow State University, Leninskie Gory, Moscow 119991, Russia}

\date{\today}
	
\begin{abstract}
Electro-optical transducers are utilized for upconvertion of radio frequency (RF) signals to the optical frequency domain to study the RF signals with optical tools. The transducers frequently impact on the RF system and introduce additional noise, including optical shot noise as well as quantum back action noise, limiting the measurement accuracy.  In this paper we theoretically study a technique based on a high efficiency electro-optical phase modulation effect that allows back action evading detection of an RF field quadrature. The main idea of proposed method is independent homodyne detection of {\em two (Stokes and anti-Stokes)} optical modes. It allows subsequent postprocessing of the detected signals, which, in turn,  results in a broadband back action evading measurement leading to the high sensitivity evaluation of the RF signals.
\end{abstract}

\maketitle

\section{Introduction}

Photonics is instrumental for generation \cite{wu96apl,lee00apl,preu11jap}, synthesis \cite{matsko19ptl}, detection and distribution \cite{lambert20aqt,huang21ao} of RF, microwave and millimeter wave signals. Significant attention has been attracted recently to the nonlinear cavity-enhanced microwave photonic devices utilized for the purpose \cite{matsko08pra,burgess09oe,hashemi09pra,burgess09oe1,strekalov09ol}. The monolithic optical microcavities, on one hand, enable efficient phase matching of the nonlinear conversion processes due to the small cavity size and, on the other hand, enable high conversion efficiency due to the high quality ($Q$-) factors. 

Quantum operation of the electro-optic transducers became feasible because of the high efficiency of the resonant electro-optical processes \cite{lambert20aqt,strekalov16jo,rueda16o,botello18o}, availability of room temperature single photon detectors as well as quantum processors operating in the RF domain. The quantum electro-optical transducers provide a way of low loss transfer and detection of the microwave quantum states \cite{matsko08pra,rueda16o,fan18sa}. 

The electro-optical transduction involves the linear electro-optic effect which corresponds to the dependence of the index of refraction of some optical crystals on a electric field amplitude of the RF signal of interest. The effect leads to the phase sensitive modulation of light. The electro-optical modulation effect was noticed as a process physically similar to the optical rectification \cite{bass62prl,armstrong62pr,bass65pr,brienza68pla}. The effect was utilized for electro-optical modulation at microwave frequencies more than a half century ago \cite{harris62apl}. The modulators constitute one of the building blocks of RF photonics, as can be seen from the multiple detailed reviews \cite{seeds06jlt, capmany07np,yao09jlt,marpaung13lpr,marpaung19np}. 

Reported previously methods assume that there is a single parametric channel transferring information from an RF photon to an optical photon. An ultimate high fidelity parametric transducer that converts RF photons to the optical frequency domain enables counting the photons with an accuracy limited by the RF photon shot noise. However, the single RF photon counters were not realized experimentally in a devices based on monolithic microcavities because of technical restrictions.

In this paper we introduce a back-action evading technique for detection of RF signals using a simple low loss resonant phase modulator capable of producing multiple optical harmonics when an RF signal is present. The upconversion of the RF signal to optical domain corresponds to establishing a photonic gain in the system \cite{matsko08pra}. For instance, if an RF photon with carrier frequency $\omega_{RF}$ is upconverted to optical frequency $\omega_0$, the power of the RF source is increased by $\omega_0/\omega_{RF}$. We show that unlike early electro-optical transducers, the device based on the phase modulation allows generation of a few optical photons per one RF photons which further increases the photonic gain in the device. This parametric amplification is accompanied by a back-action noise that masks the signal. We found that the back action can be removed from the measured signal due to quantum correlation among the optical modulation harmonics.  

\section{Model and basic equations}

\begin{figure}
    \includegraphics[width=0.45\textwidth]{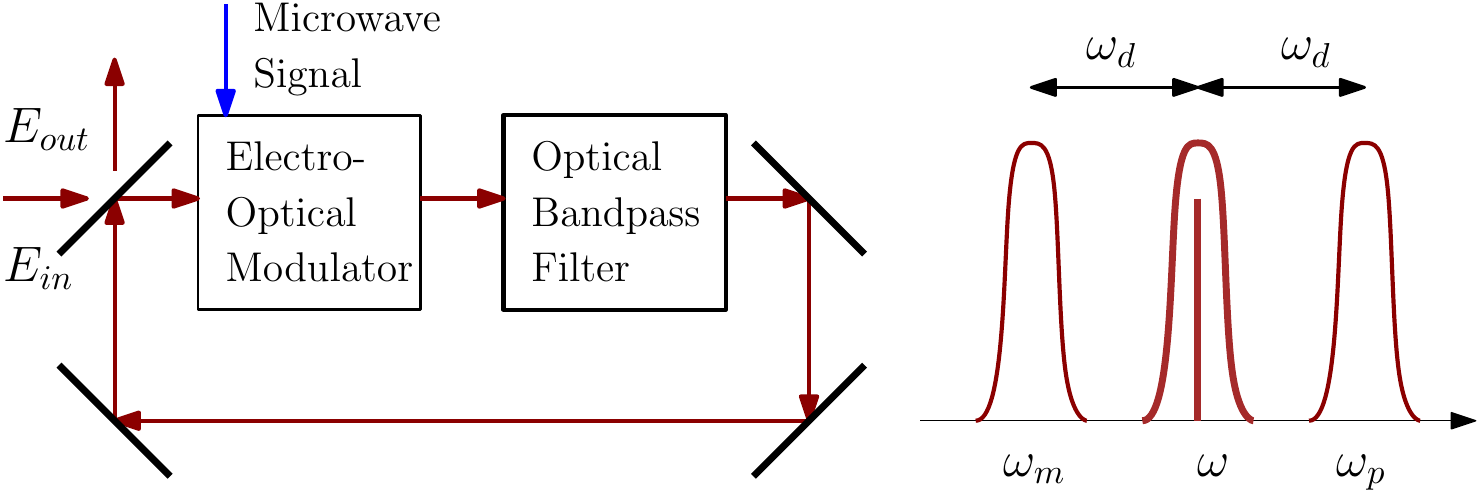}
    \caption{ Physical model of the phase-modulation based quantum electro-optical transducer. The electro-optical phase modulating crystal is inserted into an optical cavity supporting three optical modes only. This feature can be supported by insertion of a low loss optical filter into the optical cavity.  }\label{EOscheme}
   \end{figure}

Let us consider physical system depicted in Figure~(\ref{EOscheme}). An anti-reflection coated electro-optical crystal is inserted into an optical resonator supporting three optical modes only. The frequency difference between each consecutive pair of modes corresponds to the frequency of the RF signal fed into the electro-optical crystal. The central optical mode is pumped with monochromatic light.  As the result of the electro-optical interaction the light exiting the optical resonator becomes modulated. The modulation harmonics are generated in the corresponding optical modes of the resonator. To perform the measurements of the RF signal entering the electro-optical crystal we independently detect and analyze the modulation harmonics leaving the resonator. The goal of the measurement is to achieve as high as possible measurement sensitivity of the amplitude and phase components of the RF field. The three-mode configuration is introduced for the sake of simplicity of the description of the proposed measurement technique.

Let us start from the Hamiltonian of the system. We consider three optical modes including central one, described with annihilation operator $c$ and carrier frequency $\omega$, red shifted one, described with annihilation operator $c_m$ and carrier frequency $\omega_m$, and blue shifted one, described with annihilation operator $c_p$ and carrier frequency $\omega_p$.  The modes are coupling through the RF field described by annihilation operator $d$ with carrier frequency $\omega_d$. 
\begin{subequations}
\begin{align} 
H&=H_0+V,\\
H_0&=\hbar \omega c^\dag c+\hbar \omega_m c_m^\dag c_m +\hbar \omega_p c_p^\dag c_p+\hbar \omega_d d^\dag d,\\
V&=\hbar \eta(c_m^\dag d^\dag +c_p^\dag d)c+\hbar \eta^*(c_m  d  +c_p d^\dag)c^\dag.
\end{align}  
  \end{subequations} 
For the sake of simplicity we assume that the optical frequencies are equidistant and their frequency difference is equal to the frequency of the RF cavity:
\begin{equation}
\omega_p-\omega=\omega-\omega_m=\omega_d. \label{freq}
\end{equation}  
The coupling constant is measured in frequency units and is defined as 
\begin{equation} \label{eta}
\eta = \frac 12 \omega n_c^2 r \sqrt{\frac{2\pi \hbar
\omega_d}{n_d^2 {\cal V}_d}} \xi,
\end{equation}
where $n_c$ and $n_d$ are the refractive indexes of the optical and RF modes, correspondingly, $r$ is the electro-optic coefficient of the material for the selected polarizations of the modes, ${\cal V}_d$ is the volume of the RF mode,
\begin{equation}
\xi =  \frac{1}{{\cal V}}\int
\limits_{{\cal V}} d{\cal V} \Psi \Psi_{m,p} \Psi_{d}
\end{equation}
is the overlap integral of the optical and RF fields,
$|\Psi|^2$, $|\Psi_{m,p}|^2$, and $|\Psi_d|^2$ are the spatial
distributions of the power of the optical and microwave fields
respectively, $(1/{\cal V})\int |\Psi|^2 dV = 1$, ${\cal V}$ is
the mode volume for the optical cavity. We assume that the
optical modes are nearly identical, i.e. $\omega \simeq
\omega_{p,m} \gg \omega_d$, and ${\cal V} \simeq {\cal V}_{p,m}$, $\int
d{\cal V} \Psi \Psi_{p} \Psi_d = \int d{\cal V} \Psi \Psi_{m}
\Psi_d$.

Using the Hamiltonian we derive the equations for the fields, assuming all the resonant coupling. We also introduce the attenuation constants and external pumping, following the general rules of the Langevin equation formalism
\begin{subequations}
\begin{align} 
\dot c+\gamma c&= -i \eta^*(c_m  d  +c_p d^\dag)+\sqrt{2 \gamma}a,\\
\dot c_m+\gamma c_m&= -i \eta d^\dag  c+\sqrt{2 \gamma}a_m,\\
\dot c_p+\gamma c_p&=- i \eta d c +\sqrt{2 \gamma}a_p,\\
\dot d+\gamma_M d&= -i \eta c_m^\dag c -i \eta^* c_p c^\dag+\sqrt{2 \gamma_M}q +if_s.
\end{align}  \label{moveq2}
  \end{subequations} 
The external signal can be presented in form
\begin{equation}
    f_s=e^{-i\phi_s} \sqrt{\frac{2P_d\gamma_M }{\hbar \omega_d}}, \label{fs}
\end{equation}
where $P_d$ is the power of the RF signal and $\phi_s$ is the phase of the RF signal. The set of equations has to be supplied with expressions for the fields leaving the optical resonator:
\begin{equation}
b_{p,m}=-a_{p,m}+\sqrt{2 \gamma} c_{p,m}. \label{out}
\end{equation}   
It is important to note that the signal depends on the attenuation coefficient $\gamma_M$. We cannot consider a lossless mechanical system to perform the measurements discussed here.

\section{Solution of the Langevin equations}

The steady state equations for the expectation values of the field amplitudes in the case of the all-resonant tuning can be written as
\begin{subequations}
\begin{align} 
 \gamma C&= -i \eta^*(C_m  D  +C_p D^* )+\sqrt{2 \gamma}A,\\
\gamma C_m&= -i \eta D^*  C ,\\
\gamma C_p&=- i \eta DC  ,\\
\gamma_M D&= -i \eta C_m^* C -i \eta^* C_p C^* .
\end{align}  
  \end{subequations} 
Let us assume that the RF signal is weak ($D=0$). In this case $C=A\sqrt{2/\gamma}$, and $C_m=C_p=0$. 

Let us show that the solution is stable. We write a set of the equations neglecting by the external pumping:
\begin{subequations}
\begin{align} 
\dot c+\gamma c&= 0 \label{eq1},\\
\dot c_m+\gamma c_m&= -i \eta d^* C,\\
\dot c_p+\gamma c_p&=- i \eta d C,\\
\dot d+\gamma_M d&= -i  \eta c_m^* C -i \eta^* c_pC^*.
\end{align}  
  \end{subequations} 
Separating Eq.~\eqref{eq1} from the set we obtain the characteristic equation
\begin{equation}
\begin{vmatrix}
\lambda+\gamma  & 0  &  i \eta^*  C^* \\
0  & \lambda+\gamma   &   i \eta  C  \\
  i   \eta C & i \eta^* C^* & \lambda+\gamma_M
\end{vmatrix}
 =0
\end{equation}
which has all negative roots $\lambda_{1,2}=-\gamma<0$, $\lambda_3=-\gamma_M<0$, confirming that the solution is stable. 

The set of equations for the Fourier amplitudes of the fluctuation parts of the operators can be written as
\begin{subequations}
\begin{align} 
(\gamma-i \Omega) c_m^\dag&= i \eta^*d   C^*+\sqrt{2 \gamma}a_m^\dag,\\
(\gamma-i \Omega) c_p&=- i \eta d C +\sqrt{2 \gamma}a_p,\\
(\gamma_M-i \Omega)  d&= -i  \eta c_m^\dag C -i \eta^* c_p C^*+\sqrt{2 \gamma_M}q+i f_s ,\nonumber
\end{align}  \label{moveq3}
where $c_p\equiv c_p(\Omega),\ c_p^\dag \equiv c_p^\dag(-\Omega)$ and so on.

  \end{subequations} 
Introducing quadrature amplitudes
\begin{subequations}
\begin{align}
c_{pa}&=\frac{c_pe^{i \phi}+c_p^\dag e^{-i \phi}}{\sqrt{2}}, \; c_{p \phi}=\frac{c_pe^{i \phi}-c_p^\dag e^{-i \phi}}{i \sqrt{2}},\\
c_{ma}&=\frac{c_me^{i \phi}+c_m^\dag e^{-i \phi}}{\sqrt{2}}, \; c_{m \phi}=\frac{c_me^{i \phi}-c_m^\dag e^{-i \phi}}{i \sqrt{2}},\\
d_a &= \frac{d+d^\dag}{\sqrt 2},\quad d_\phi = \frac{d-d^\dag}{i\sqrt 2}.
\end{align}  
  \end{subequations} 
where $\eta C= |\eta C| e^{-i \phi}$. Using Eq.~\eqref{moveq3} we derive
\begin{subequations}
\begin{align}
(\gamma-i \Omega) c_{ma}&=-|\eta C|d_\phi+\sqrt{2 \gamma}a_{ma},\\
(\gamma-i \Omega) c_{m \phi}&=-|\eta C|d_a+\sqrt{2 \gamma}a_{m \phi},\\
(\gamma-i \Omega) c_{pa}&=|\eta C|d_\phi+\sqrt{2 \gamma}a_{pa},\\
(\gamma-i \Omega) c_{p \phi}&=-|\eta C|d_a+\sqrt{2 \gamma}a_{m \phi},\\
(\gamma_M-i \Omega)  d_a&= -| \eta C| c_{m \phi}+| \eta C| c_{p \phi}+\sqrt{2 \gamma_M}q_a-f_\phi,\nonumber\\
(\gamma_M-i \Omega)  d_\phi&= -| \eta C| c_{m a}-| \eta C| c_{p a}+\sqrt{2 \gamma_M}q_\phi+f_a.\nonumber
\end{align}   \label{quad1}
  \end{subequations} 
Let us assume that we are able to measure output waves from modes $c_m$ and $c_p$ separately. This is reasonable since the microwave frequency can exceed tens of GHz and low loss filters with narrower bandwidth exists. Let us introduce linear combinations of the quadrature amplitudes
\begin{subequations}
\begin{align} 
\alpha_{\pm a}=\frac{a_{pa}\pm a_{ma}}{\sqrt 2}, \qquad \alpha_{\pm \phi}=\frac{a_{p \phi}\pm a_{m \phi}}{\sqrt 2},\\
g_{\pm a}=\frac{c_{pa}\pm c_{ma}}{\sqrt 2}, \qquad g_{\pm \phi}=\frac{c_{p \phi}\pm c_{m \phi}}{\sqrt 2},\\
\beta_{\pm a}=\frac{b_{pa}\pm b_{ma}}{\sqrt 2}, \qquad \beta_{\pm \phi}=\frac{b_{p \phi}\pm b_{m \phi}}{\sqrt 2}.
\end{align}  
  \end{subequations} 
and rewrite set \eqref{quad1} in form
\begin{subequations}
\begin{align}
(\gamma-i \Omega) g_{+a}&= \sqrt{2 \gamma}\alpha_{+a},\\
(\gamma-i \Omega)g_{-a}&=\sqrt{2}|\eta C|d_\phi+\sqrt{2 \gamma}\alpha_{-a},\\
\label{dphi}
(\gamma_M-i \Omega)  d_\phi&= -\sqrt{2}| \eta C|g_{+ a} +\sqrt{2 \gamma_M}q_\phi+f_a,\\
(\gamma-i \Omega) g_{+ \phi}&=-\sqrt{2}|\eta C|d_a+\sqrt{2 \gamma}\alpha_{+ \phi},\\
(\gamma-i \Omega) g_{-\phi}&= \sqrt{2 \gamma}\alpha_{- \phi},\\
(\gamma_M-i \Omega)  d_a&= \sqrt{2}| \eta C| g_{- \phi}+\sqrt{2 \gamma_M}q_a-f_\phi.
\end{align}  
  \end{subequations} 
In addition, for the output fields we get using \eqref{out}
\begin{subequations}
\begin{align}
\label{beta+a}
\beta_{+a} &=-\alpha_{+a}+\sqrt{2 \gamma} g_{+a}=\frac{\gamma+i \Omega}{\gamma-i \Omega}\alpha_{+a},\\
\label{beta-a}
\beta_{-a} &=\frac{\gamma+i \Omega}{\gamma-i \Omega}\alpha_{-a}+\frac{2\sqrt{\gamma}|\eta C|}{\gamma-i \Omega}d_{\phi},\\
\beta_{-\phi}&=-\alpha_{-\phi}+\sqrt{2 \gamma} g_{-\phi}=\frac{\gamma+i \Omega}{\gamma-i \Omega}\alpha_{-\phi},\\
\beta_{+\phi}&= \frac{\gamma+i \Omega}{\gamma-i \Omega}\alpha_{+\phi}-\frac{2\sqrt{\gamma}|\eta C|}{\gamma-i \Omega}d_{a}.
\end{align}   \label{bb}
  \end{subequations} 

Introducing notations
\begin{subequations}
\begin{align}
\beta=\frac{\gamma+i \Omega}{\gamma-i \Omega},\quad
K=\frac{4\gamma|\eta C|^2}{\gamma^2+\Omega^2}.
\end{align}
\end{subequations}
we rewrite Eqs.~\eqref{bb} as
\begin{subequations}
\begin{eqnarray}
\beta_{-a}
=\beta\alpha_{-a}-\frac{\beta K}{ (\gamma_M-i \Omega)}\alpha_{+a}+\\ \nonumber \frac{\sqrt{\beta K} }{(\gamma_M-i \Omega)}(\sqrt{2 \gamma_M}q_\phi+f_a),\\
\beta_{+\phi}
=\beta\alpha_{+\phi}-\frac{\beta K}{ (\gamma_M-i \Omega)}\alpha_{-\phi}-\\ \nonumber \frac{\sqrt{\beta K}}{(\gamma_M-i \Omega)}(\sqrt{2 \gamma_M}q_a-f_\phi)
\end{eqnarray}
\label{betas}
  \end{subequations} 
Here the terms proportional to $K$ result from the back action. 

\section{Measurement strategy}
We see that sum $\beta_{+a}$ \eqref{beta+a} of  output quadratures contains only input fluctuations $\alpha_{+a}$, whereas difference $\beta_{-a}$ \eqref{beta-a} contains both input fluctuations $\alpha_{-a}$ and term proportional to RF phase quadrature  $d_\phi$. In turn, $d_\phi$ \eqref{dphi} depends on back action $\alpha_{+a}$. Hence, post processing of independent measurements of $\beta_{+a}$ and $\beta_{-a}$ gives us possibility of {\em complete} exclusion of back action. For this we have to take  the following combinations
\begin{subequations}
 \begin{equation} \label{ampnoise}
\begin{split}
\beta_a=\frac{  K}{ (\gamma_M-i \Omega)}\beta_{+a}+\beta_{-a}=\\
=\beta\alpha_{-a}+\frac{\sqrt{\beta K}}{ (\gamma_M-i \Omega)}(\sqrt{2 \gamma_M}q_\phi+f_a)
\end{split}
\end{equation}
 \begin{equation}
\begin{split}
\beta_\phi=\frac{K}{ (\gamma_M-i \Omega)}\beta_{-\phi}+\beta_{+\phi}=\\
=\beta\alpha_{+\phi}-\frac{\sqrt{\beta K}}{ (\gamma_M-i \Omega)}(\sqrt{2 \gamma_M}q_a-f_\phi)
\end{split}
\end{equation}
  \end{subequations}

In this scheme the SQL arises when we measure only one combination $\beta_{-a}$ or $\beta_{+\phi}$. For the case when we measure  $ \beta_{-a}$ \eqref{beta-a} we get
 \begin{align}
  \label{Sf}
  S_{n}(\Omega) &= 2\gamma_M\big(2n_T+1\big) 
    + \frac{\big(\gamma_M^2+\Omega^2\big)}{K} +K\ge\\
    \label{SQL}
    &\ge 2\gamma_M\big(2n_T+1\big) +S_{SQL},\\ 
\label{Ssql}
  S_{SQL} &= 2\sqrt{\gamma_M^2+\Omega^2}, \quad K_{SQL}=\sqrt{\gamma_M^2+\Omega^2}.
 \end{align}
The sensitivity is restricted by SQL. The first term $2\gamma_M(2n_T+1)$ corresponds to the RF limit \eqref{rfl}, discussed below. We should note that in this scheme even in the case of zero temperature ($n_T=0$) we can not set the attenuation coefficient $\gamma_M=0$ to fully neglect the RF fluctuations and redefine the SQL.  It is because the $\gamma_M$ serves as the coupling between the RF signal and the electro-optical crystal \eqref{fs}.

\begin{figure}
    \includegraphics[width=0.45\textwidth]{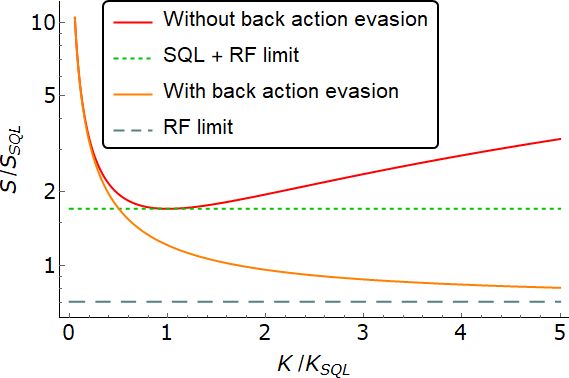}
 \caption{Noise power spectral densities $S$ as a function of power parameter $K$.   Measurement of only $\beta_{-a}$ or $\beta_{+\phi}$ \eqref{Sf} (red line) does not allow to surpass the sum of SQL and RF limit \eqref{SQL} (dashed green line). Measurement of combination $\beta_{a}$ \eqref{noise} (orange line)  allows to  evade back action and surpass SQL, but the sensitivity is limited by the RF limit \eqref{rfl} (dashed blue line).
The plots are presented for  $\Omega/2\pi=\gamma_M/2\pi=1$ MHz and $n_T=0$.}\label{plots1}
   \end{figure}

If we rewrite Eq.~(\ref{ampnoise}) in form 
\begin{eqnarray}
\beta_a=\frac{\sqrt{\beta K}}{ \gamma_M-i \Omega} \left ( \frac{ \gamma_M+i \Omega}{\sqrt{\beta K}}  \alpha_- +  \sqrt{2 \gamma_M}q_\phi+f_a  \right )
\end{eqnarray}
which results in the normalized spectral power density of the quantum noise.
In this case the detection condition becomes
\begin{align}
 \label{noise}
    S_{n}(\Omega) & = 2\gamma_M\big(2n_T+1\big) + \frac{\gamma_M^2 +\Omega^2 }{ K}
\end{align}
Here we assume that the single sided spectral density of  $\alpha_{a-}$ is equal $1$.  The power spectral densities are illustrated by Fig.~\ref{plots1}.

The detection condition
\begin{equation}
   \label{fso}
    f_{s0}\ge \sqrt {S_{n}(\Omega)\, \frac{\Delta\Omega}{2\pi}}
\end{equation}
indicates that the suggested here measurement procedure leads to a broadband force measurement with sensitivity better than the SQL. We note that the external signal is defined in a way that $f_{s0}$ is a generalized force acting on the RF oscillator. It can be recounted into the actual RF power or the number of RF photons using the definition \eqref{fs}.

\begin{figure}
    \includegraphics[width=0.45\textwidth]{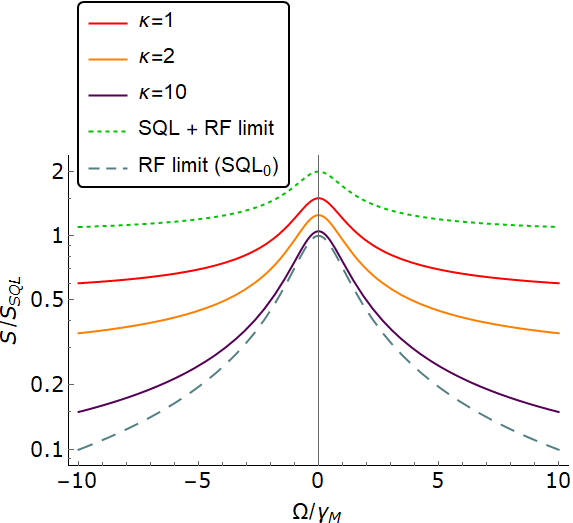}
 \caption{Noise power spectral densities $S$ as a function of frequency $\Omega$ for different power parameters $\kappa=K/K_{SQL}$.   Measurement of combination $\beta_{a}$ \eqref{noise} (red, orange and purple lines)  allows to  evade back action and surpass the sum of SQL and RF limit \eqref{SQL}  (dashed green line) for $\kappa>0.5$, but the sensitivity is limited by the RF limit \eqref{rfl} (dashed blue line).   
 The plots are presented for  $\gamma_M/2\pi=1$ MHz and $n_T=0$.}\label{plots2}
   \end{figure}

However, we should note that in the case of zero temperature ($n_T=0$) and resonant signal $(\Omega=0$) the SQL \eqref{Ssql} is equal to the RF limit \eqref{rfl} $S_{SQL_0}=S_{SQL}(0)=S_{RF}=2\gamma_M$. So even though we can beat the SQL when measuring the other spectral components of the signal,  we can only reach the SQL for the resonant component, as it can be seen on Fig.~\ref{plots2}.

\begin{figure}
    \includegraphics[width=0.445\textwidth]{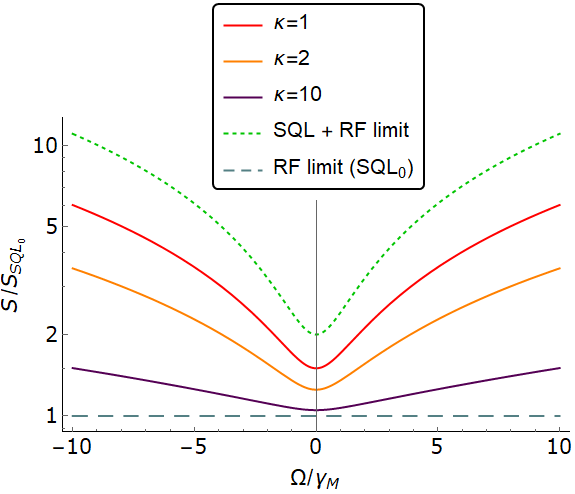}
 \caption{The same curves  as on Fig.~\ref{plots2} but the power spectral density is normalized to $S_{SQL}(0)$ to illustrate that we can not beat it and offresonant sensitivity is worse than resonant. 
 The plots are presented for  $\gamma_M/2\pi=1$ MHz and $n_T=0$.}\label{plots3}
   \end{figure}

Eq. \eqref{Ssql} shows that the SQL gets higher with the increase of $\Omega$. This is the reason why we can beat the SQL at the nonresonant frequencies. It would be more natural to watch the value $S/S_{SQL_0}$. If we compare the sensitivity in a such way, we can say that it actually gets worse when we measure the offresonant components. It is presented on Fig.~\ref{plots3} (the plots on Fig.~\ref{plots1} will not change because they are plotted for the given value of $\Omega$).

The back action is excluded completely in output quadrature $\beta_a$. This is a consequence of the separate measurements of the correlated output waves escaping modes $c_m$ and $c_p$. This is the main result of our paper. 


\section{Discussion}

It is interesting to note than the proposed physical system allows achieving high measurement sensitivity since it allows generation of multiple optical photons per single microwave photons. Indeed, for a relatively strong RF signal the averaged number of photons in an optical sideband is given by
\begin{equation}
    |C_{p,m}|^2= \frac{K}{2\gamma} \frac{P_d}{\hbar \omega_d \gamma_M} \approx \frac{K}{4\gamma} |D|^2,
\end{equation}
where $|D|^2$ is the average number of photons entering the resonator during the ringdown time and $|C_{p,m}|^2$ is the average number of photons leaving the resonator during the same time. Assuming that $K/(4 \gamma) >1$, we conclude that a single RF photon in a cavity can generate more than one optical photon. This is the reason for the high measurement sensitivity with the two output channels. 

This is different from the behavior of the transducer having only one harmonic \cite{matsko08pra}. In that receiver the RF photon can be converted to optical photon with unity probability. 

Let us assume that the mode $c_m$ is absent,  then we get
\begin{subequations}
\begin{align} 
\dot c_p+\gamma c_p&=- i \eta d C,\\
\dot d+\gamma_M d&= -i \eta^* c_pC^*.
\end{align}  
  \end{subequations} 
In the frequency domain the set is transferred to
\begin{align} 
(\gamma-i \Omega) c_p + i \eta  C d &=\sqrt{2 \gamma}a_p,\\
i \eta^*  C^* c_p +(\gamma_M-i \Omega)  d&= \sqrt{2 \gamma_M}q+i f_s ,
\end{align} 
and the solution of the set can be presented in form
\begin{align} 
 c_p &=\frac{(\gamma_M-i \Omega)\sqrt{2 \gamma}a_p 
	 -i\eta C \left(\sqrt{2 \gamma_M}q+i f_s\right)}{(\gamma-i \Omega)(\gamma_M-i \Omega) + |\eta|^2|C|^2}
\end{align}

The output amplitude can be found using \eqref{out}
\begin{subequations}
\begin{align}
 b_p &=-a_p +\sqrt{2\gamma}\, c_p =\\ \label{ba}
 &= \frac{\left[(\gamma_M-i \Omega)(\gamma + i\Omega)- |\eta C|^2\right] a_p }{(\gamma-i \Omega)(\gamma_M-i\Omega)
  +|\eta C|^2}+\\
&\quad +\frac{ -i\eta C\sqrt{2\gamma} \left(\sqrt{2 \gamma_M}q+i f_s\right)}{(\gamma-i \Omega)(\gamma_M-i \Omega)
  + |\eta C|^2} \label{sig}
\end{align}
\end{subequations} 
We find that with increase of the pump ($C$) the magnitude of the term containing the signal (\ref{sig}) does not increase and there is an optimum pump at which the back action term (\ref{ba}) is small.

Let analyse a particular realization of the scheme in which the relaxation rate of the RF mode is much smaller than relaxation rate of optical mode ($\gamma_M \ll \gamma$). The optimal pump power is given by
  \begin{align}
   \gamma_M = \frac{|\eta C|^2}{\gamma} 
  \end{align}

In this case the contribution of the shot noise descrived by Eq.~(\ref{ba}) can be dropped and the detection condition (\ref{fso}) is defined by the noise spectral density
 \begin{align}
S_{RF}  = 2\gamma_M\big(2n_T+1\big) \label{rfl}
 \end{align}
It means that the sensitivity of the measurement involving the single optical sideband is defined by the RF fluctuations only. The optical fluctuations practically do not contribute to the measurement noise. To achieve similar conditions in the case considered in this paper we have to use high enough pump power ($K\gg 1$). 

The power of the signal in the single sideband case is limited by the input RF power. The optical harmonic power is $P_d (\omega/\omega_d)^2$. In our case the power of each harmonic is  $P_d (\omega/\omega_d)^2 (K/4\gamma)$. Therefore, even though the sensitivity in both measurement schemes is similar, our system is capable of producing higher power signals being easier to detect.

\section*{Acknowledgments}
The research of SPV and AIN has been supported by the Russian Foundation for Basic Research (Grant No. 19-29-11003), the Interdisciplinary Scientific and Educational School of Moscow University ``Fundamental and Applied Space Research'' and from the TAPIR GIFT MSU Support of the California Institute of Technology. AIN is the recipient of a Theoretical Physics and Mathematics Advancement Foundation “BASIS” scholarship (Contract No. 21-2-10-47-1). The reported here research performed by ABM was carried out at the Jet Propulsion Laboratory, California Institute of Technology, under a contract with the National Aeronautics and Space Administration (80NM0018D0004).

\end{document}